\documentclass[aip,amsmath,amssymb,reprint,]{revtex4-1}
\usepackage{amsmath}
\usepackage{cancel}
\usepackage{braket}
\usepackage{amssymb}
\usepackage{color}
\usepackage{graphicx}
\usepackage{tabularx}
\usepackage{dcolumn} 
\usepackage{bm}      
\usepackage{array}   
\usepackage{amsfonts}
\usepackage{filecontents}
\usepackage{url}
\usepackage{xr-hyper}
\usepackage[bookmarks=false,colorlinks,citecolor=red]{hyperref}
\usepackage[utf8]{inputenc}
\usepackage[T1]{fontenc}
\usepackage{mathptmx}

\makeatletter
\newcommand*{\addFileDependency}[1]{
  \typeout{(#1)}
  \@addtofilelist{#1}
  \IfFileExists{#1}{}{\typeout{No file #1.}}
}
\makeatother

\newcommand*{\myexternaldocument}[1]{
    \externaldocument{#1}
    \addFileDependency{#1.tex}
    \addFileDependency{#1.aux}
}
\myexternaldocument{supplement}

\newcommand{\DSS}[1]{{\color{red}{#1}}}

\begin{document}
\include{MyCommand}
\title{A first-principles study of bilayer 1T'-WTe$_2$/CrI$_3$: A candidate topological spin filter}

\author{Daniel Staros}
\affiliation{Department of Chemistry, Brown University, Providence, RI 02912}
\author{Brenda Rubenstein}
\email{Authors to whom correspondence should be addressed: Brenda Rubenstein, brenda\_rubenstein@brown.edu, Daniel Staros, daniel\_staros@brown.edu, and Panchapakesan Ganesh, ganeshp@ornl.gov}
\affiliation{Department of Chemistry, Brown University, Providence, RI 02912}
\author{Panchapakesan Ganesh}
\affiliation{Center for Nanophase Materials Sciences, Oak Ridge National Laboratory, Oak Ridge, TN 37831, USA}

\date{\today}

\begin{abstract}
The ability to manipulate electronic spin channels in 2D materials is crucial for realizing next-generation spintronics. Spin filters are spintronic components that polarize spins using external electromagnetic fields or intrinsic material properties like magnetism. Recently, topological protection from backscattering has emerged as an enticing feature that can be leveraged to enhance the robustness of 2D spin filters. In this work, we propose and then characterize one of the first 2D topological spin filters: bilayer CrI$_3$/1$T$'-WTe$_2$. To do so, we use a combination of Density Functional Theory, maximally localized Wannier functions, and quantum transport calculations to demonstrate that a terraced bilayer satisfies the principal criteria for being a topological spin filter; namely that it is gapless,
exhibits spin-polarized charge transfer from WTe$_2$ to CrI$_3$ that renders the bilayer metallic, and has a topological boundary which retains the edge conductance of monolayer 1$T$'-WTe$_2$. In particular, we observe that small negative ferromagnetic moments are induced on the W atoms in the bilayer, and the atomic magnetic moments on the Cr are approximately 3.2 $\mu_B$/Cr  compared to 2.9 $\mu_B$/Cr in freestanding monolayer CrI$_3$. Subtracting the charge and spin densities of the constituent monolayers from those of the bilayer further reveals spin-orbit coupling-enhanced spin-polarized charge transfer from WTe$_2$ to CrI$_3$. We find that the bilayer is topologically trivial by showing that its Chern number is zero. Lastly, we show that interfacial scattering at the boundary between the terraced materials does not remove WTe$_2$'s edge conductance. Altogether, this evidence indicates that BL 1$T$'-WTe$_2$/CrI$_3$ is gapless, magnetic, and topologically trivial, meaning that a terraced WTe$_2$/CrI$_3$ bilayer heterostructure in which only a portion of a WTe$_2$ monolayer is topped with CrI$_3$ is a promising candidate for a 2D topological spin filter. Our results further suggest that 1D chiral edge states may be realized by stacking strongly ferromagnetic monolayers, like CrI$_3$, atop 2D $non$magnetic Weyl semimetals like 1$T$'-WTe$_2$. 
\end{abstract}

\maketitle

\pagenumbering{arabic}

\section*{Introduction}
\noindent
As signs continue to suggest that Moore's Law
has plateaued, researchers have begun to seek new routes to designing faster, smaller, more energy-efficient, and more versatile electronic devices. The key to realizing such devices will be discovering, characterizing, and designing novel nanoscale quantum electronic components whose many electronic degrees of freedom, including their electron spin\cite{Wolf_science2001,Ahn_npj2020} and momenta,\cite{zhang_two-dimensional_2021,zhang_2d_2022} can be manipulated to enable faster, more energy-efficient operations on denser data. 

Along these lines, nanoscale spintronics have been hailed as extremely promising routes towards denser data storage and potentially faster and more efficient reading and writing. Unlike conventional electronics which harness the charge of an electron, spintronic materials store information in electrons' two possible spin states,\cite{Wolf_science2001} which can be manipulated more rapidly and with less energy than electrons' charges.\cite{ElGhazaly_jmmm2020} Spintronic devices are also less volatile than conventional electronic devices because they can preserve their spin even in the absence of electric power.\cite{Wolf_science2001,Barla_jce2021} Moreover, one of the primary advantages of spintronic devices is that they can be readily integrated into modern CMOS-based circuits.\cite{Barla_jce2021}

Since the birth of spintronics with the discovery of the giant magnetoresistive effect,\cite{Wolf_science2001} the world of spintronic device components has expanded to include various spintronic analogues to traditional resistors and transistors, as well as new components unique to controlling spin currents like spin filters and spin injectors.\cite{Hirohata_jmmm2020} Many of these components take advantage of the properties of magnetic materials, in which spins are already selectively ordered. For example, two spintronic analogues to traditional resistors, spin valves and magnetic tunnel junctions (MTJs), typically consist of two ferromagnetic layers separated by an insulating layer.\cite{Barla_jce2021,Hirohata_jmmm2020} Varying the magnetic orientation of one of the magnetic layers and keeping the other fixed allows the resistance to spin currents to be changed by taking advantage of spin-selective quantum tunneling as in MTJs, or the giant magnetoresistive or spin-transfer torque effects as in spin valves. Perhaps the most fundamental spintronic device component, however, is that which enables the generation of spin current in the first place: the spin filter.

\begin{figure}[ht!]
  \setlength{\belowcaptionskip}{-16pt}
   \setlength{\abovecaptionskip}{0pt}
  \begin{center}
    \includegraphics[width = 1\linewidth]{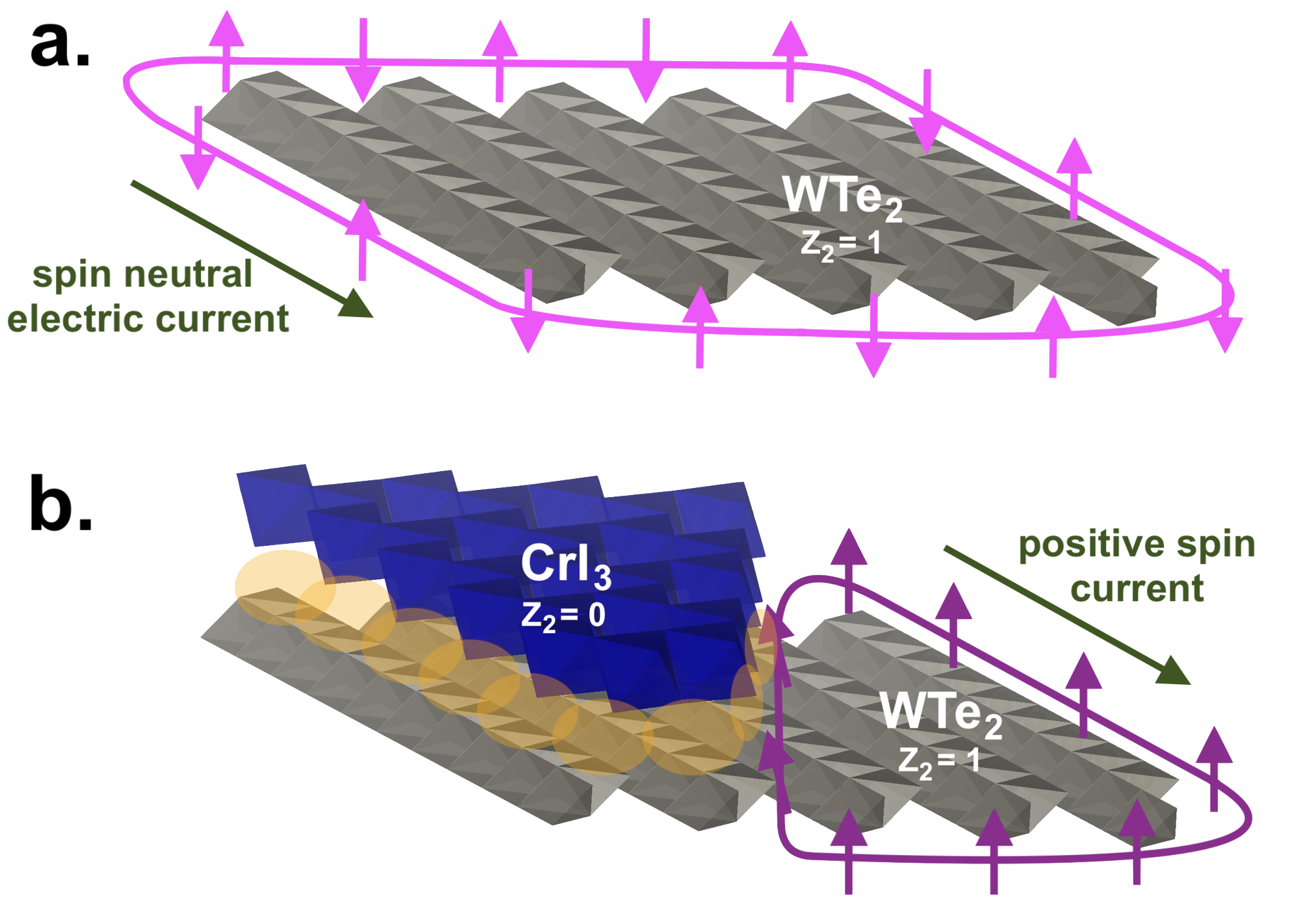}
  \caption{(a) Edge conductance around ML WTe$_2$. (b) Proposed spin-polarized edge conductance/1D chiral edge state of the CrI$_3$/WTe$_2$ bilayer studied here.} 
  \label{fig1}
  \end{center}
\end{figure}
\noindent

Spin filters are devices that generate spin-polarized currents from unpolarized electric currents by selectively transmitting electrons with a particular spin and blocking those with the opposite spin. In general, such devices have most often taken advantage of the inherent spin polarization in ferromagnetic or multiferroic materials,\cite{Inomata_stam2008,Bea_jpcm2008,Zhang_npjcm2022} spin-selective quantum tunneling using barrier materials with different spin-dependent transmission probabilities,\cite{Massarotti_natcomm2015} or spin orbit coupling in Rashba-type spin filters to achieve this.\cite{Koga_prl2002,Cummings_apl2006} Spin filters can additionally be designed by placing ferromagnetic insulators in close proximity to superconducting junctions since the presence of an out-of-plane magnetic field can break time-reversal symmetry, causing the electrons to form spin-polarized currents through the spin-transfer torque effect.\cite{Hasan_rmp2010,Mak_natrev2019,Yoshida_jpc2019} 

Recently, the concept of topological spin filters has been put forth as one promising option for improving the robustness of spin filters at higher temperatures by taking advantage of quantum anomalous Hall conductance, which is topologically protected from backscattering and could minimize dissipation as a result. By extension, edge-conductance dominated by one spin channel amounts to nearly dissipationless conductance of one direction of quantum spin, also known as topological spin filtering, which can manifest as a chiral edge state along the edge of a partially exposed topologically trivial bilayer and nontrivial monolayer (see Figure \ref{fig1}).\cite{Howard_natcomm2021} The magnetic Weyl semimetal Co$_3$Sn$_2$S$_2$ has recently been discussed as a potential avenue towards realizing such higher temperature chiral conducting edge states,\cite{Hsieh_prl2016, Howard_natcomm2021} which could in principle also become spin-polarized. Along the same vein, a topological spin filter may be constructed by placing a ferromagnet near \textit{topologically nontrivial} 2D materials such as 1$T'$ transition metal dichalcogenides,\cite{Qian_science2014} which could also give rise not only to spin-polarized currents, but spin-polarized helical edge modes that are topologically protected from backscattering.

While many such three-dimensional spin filters have been proposed, two-dimensional materials and their heterostructures possess a larger design space advantageous for engineering new spintronic devices.\cite{Sierra_natnano2021} Two-dimensional heterostructures can be designed to exhibit a wide array of emergent properties by mixing and matching the properties of their constituent monolayers,\cite{Novoselov_science2016,Pham_chemrev2022} twisting them,\cite{he_moire_2021,xu_coexisting_2022, Morrissette_natphys2023, fumega_moire-driven_2023} straining them,\cite{dai_strain_2019} or placing them in proximity to electric and/or magnetic fields.\cite{Maximenko_npj2022,Marmodoro_prb2022,tao_multiband_2022,iordanidou_electric_2023} Despite this, researchers have only recently made significant strides towards truly 2D spin filters which promise to be smaller, more tunable, and ideally more efficient than their 3D counterparts. Graphene is one 2D material that originally garnered spintronic interest when it was predicted to exhibit nearly perfect spin filtering when interfaced with only a couple of layers of a ferromagnetic metal.\cite{Karpan_prl2007} However, experimental attempts to realize such a graphene-based spin filter fell short, initially showing tunnel magnetoresistance ratios of 0.4\% for graphene/NiFe, with additional attempts increasing this ratio to no more than 5\%.\cite{Ahn_npj2020,Piquemal_jpd2017,Cobas_nano2012}
More recently, the pivotal discovery of giant magnetoresistance in bilayer CrI$_3$ in 2018 has reinvigorated the search for better 2D spin filters based on atomically thin magnets,\cite{Song_science2018,Wang_natcomm2018,Klein_science2018,Zhu_prb2021} and a slew of inspired studies have since been published that take advantage of their properties.\cite{Zhu_prb2021, Luo_pra2023, Li_apl2023, Li_npj2023} 
These examples suggest that, with the right combination of 2D monolayers, 2D spin filters, and even 2D topological spin filters, should also be within reach. 

Notably, there are few studies which consider the proximity effects of 2D magnets stacked atop a monolayer of $1T'$-WTe$_2$, which is the only MX$_2$ monolayer that exists in the $1T'$ phase in its ground state and the only such member which is topological as a freestanding monolayer.\cite{Tang_natphys2017} Until fairly recently, the only such example consisted of one layer of $1T'$-WTe$_2$ interfaced with one layer of permalloy (Ni$_{80}$Fe$_{20}$) to form a film with several-nm thickness that exhibited out-of-plane magnetic anistropy 
.\cite{MacNeill_natPhys2017,Li_natcomm2018} More recently, proximity-induced magnetic order was observed in monolayer $1T'$-WTe$_2$ placed onto antiferromagnetic trilayer CrI$_3$, where edge conductance jumps were observed upon switching of CrI$_3$’s magnetic state.\cite{Chang_natmat2020,Zhao_natmat2020} Most recently, proximity-induced half-metallicity and complete spin-polarization was predicted in bilayer $1T'$-WTe$_2$/CrBr$_3$ and attributed to strong orbital hybridization and charge transfer at the interface of the heterostructure.\cite{Bora_ass2023} Nonetheless, to the best of our knowledge, no investigation of the topological properties of a 1$T'$-WTe$_2$/CrX$_3$ bilayer has yet been performed, let alone with the goal of realizing a new type of \textit{topological} spin filter. Taken together, these discoveries point towards bilayer 1$T'$-WTe$_2$/CrI$_3$ as a strong potential candidate for topological spin filtering which could leverage the perfect spin filtering of a 1$T'$-WTe$_2$/CrX$_3$ hetrostructure in proximity to the dissipationless edge states of $1T'$-WTe$_2$.

Thus, in this manuscript, we use \textit{ab initio} and quantum transport simulations to identify terraced bilayer 1$T'$-WTe$_2$/CrI$_3$ as a promising candidate for a 2D topological spin filter. 1$T'$-WTe$_2$ is a nonmagnetic Weyl semimetal which exhibits topological edge conductance in its monolayer form,\cite{Qian_science2014}
while monolayer CrI$_3$ is a ferromagnetic Mott insulator.\cite{Huang_natlett2017}  One can thus imagine that, by placing these two materials in proximity, the CrI$_3$'s magnetism could potentially polarize WTe$_2$'s edge conductance, forming a topological spin filter. To determine whether this is in fact the case, we predict the band structures, topological invariants, and interlayer charge and magnetization density transfer for BL CrI$_3$/1$T'$-WTe$_2$ with and without spin-orbit coupling. In so doing, we unequivocally demonstrate that the proximity of CrI$_3$ to WTe$_2$ foremost results in strong interlayer coupling between the two layers, spin-polarization on the WTe$_2$, and an overall trivial BL topology. To determine whether edge conductance is lost in a terraced bilayer, we also predict the conduction in a model terraced bilayer, showing that spin-polarized edge conductance is retained. These considerations, taken together with the metallic nature of the bilayer and previous evidence for spin-polarized helical edge modes in monolayer $1T'$-WTe$_2$, provide convincing evidence for the possibility of realizing chiral edge states at the interface of a terraced $1T'$-WTe$_2$/CrI$_3$ bilayer. Specifically, our results imply that electric current injected into the metallic bilayer portion of the terraced heterostructure would become spin-polarized before transferring to WTe$_2$ and exiting via spin-polarized edge conductance in chiral edge states around the monolayer WTe$_2$ portion; this terraced, strained $1T'$-WTe$_2$/CrI$_3$ bilayer is then likely a strong candidate for a highly robust, ultra-thin spin filter with 1D chiral edge states.

\section*{Methods \label{methods}}

In order to determine the ground electronic states of the constituent monolayers and 1$T'$-WTe$_2$/CrI$_3$ bilayer, as well as the charge transfer present in the bilayer, we calculated ground electronic states and charge and spin densities using self-consistent Density Functional Theory (DFT). We then Wannierized the DFT orbitals we obtained into a form allowing for the calculation of Chern numbers, which determine whether a material is topologically nontrivial. Finally, we characterized the conduction observed in our heterostructure by extending a previously parameterized $\textbf{k} \cdot \textbf{p}$ model of 1$T'$-WTe$_2$ to a simplified version of our terraced heterostructure. As the different rhombohedral angles of monolayer $R3$ CrI$_3$ and 1$T'$-WTe$_2$ do not lend themselves to the simple construction of a commensurate supercell without the introduction of different strain to both layers, the layers were strained slightly by hand so that they could share a common cell. We therefore begin this methodology section by discussing the determination of the appropriate strain and interlayer distance of the bilayer before going into greater detail about the DFT calculations, the computation of the Chern numbers, and the details of our edge conductance simulations.

\subsection*{Bilayer Supercell Construction}

To construct our bilayer, we used a highly-accurate DMC-optimized monolayer CrI$_3$ structure containing 8 atoms in its unit cell and exhibiting triclinic ($R3$) symmetry.\cite{Staros_jcp2022} A monolayer $1T$'-WTe$_2$ structure was obtained from the Materials Project website.\cite{Jain_aplm2013} During the simulation of these monolayers, more than 20 Å of vacuum was added to both structures to prevent spurious self interactions. 

\begin{figure*}[htp!]
  \setlength{\belowcaptionskip}{-16pt}
   \setlength{\abovecaptionskip}{0pt}
  \begin{center}
    \includegraphics[width = 6.0in]{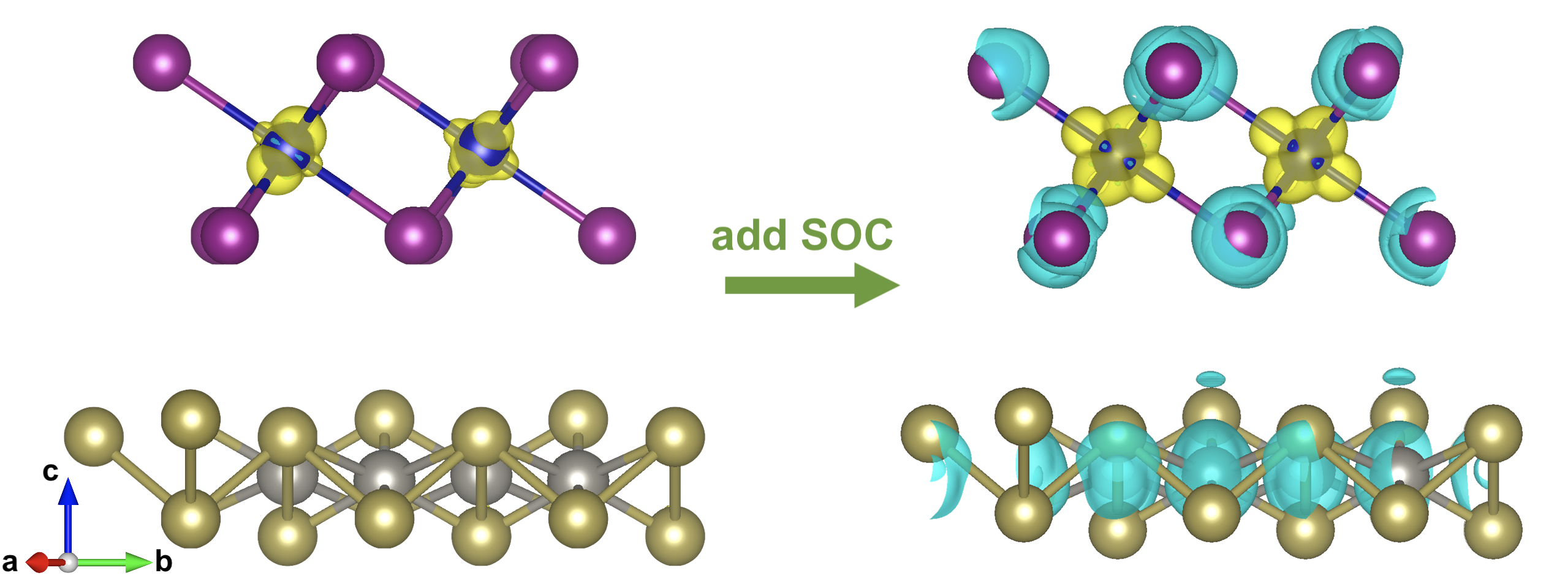}
  \caption{(Left) Collinear PBE+$U$ spin density difference (yellow = positive) $s_{BL}-s_{CrI3}-s_{WTe2}$ with an isosurface level value of 0.0025 shows slightly enhanced positive magnetization on the Cr atoms due to spin-polarized charge transfer. (Right) Noncollinear PBE+$U$ magnetization density $z$-component difference (yellow = positive) $m_{BL}-m_{CrI3}-m_{WTe2}$ with an isosurface level value of 0.0025 shows more drastic Cr magnetization enhancement, and additionally magnetic induction in WTe$_2$.} 
  \label{fig2}
  \end{center}
\end{figure*}

In stacking the layers, special consideration was given to how to align them since their monolayer structures are incommensurate. In particular, the WTe$_2$ cell was rotated such that the original lattice constants for the monolayer cells were strained as little as possible. The result of this process was a bilayer with a lattice constant of 7.01 Å, which means that our DMC-optimized ML CrI$_3$ structure is stretched by 2.5\% relative to the monolayer, and WTe$_2$ (a = 3.505 Å) is stretched by 0.8\% relative to the experimental bulk 1$T'$-WTe$_2$ value of 3.477 Å.\cite{Mar_jacs1992} Additionally, our monolayer 1$T'$-WTe$_2$ lattice constant is close to the value of 3.502 Å previously obtained using DFT structural relaxation.\cite{Xiang_aipa2016} 

\subsection*{Density Functional Simulations}

All simulations of structural and electronic properties were performed using DFT as implemented within the Quantum ESPRESSO package.\cite{Giannozzi_jpcm2009,Giannozzi_jpcm2017} The PBE and PBE+$U$ functionals\cite{perdew1996} with $U=2$ eV on the chromium atoms were selected to model these materials because previous studies demonstrated that a trial wavefunction utilizing a Hubbard $U$ value of 2-3 eV minimizes the fixed-node error in DMC calculations of CrI$_3$.\cite{Staros_jcp2022, Ichiba_prm2021} Our calculations used norm-conserving, scalar-relativistic Cr and relativistic I pseudopotentials and recently developed spin–orbit relativistic effective W and Te pseudopotentials.\cite{Wang_jcp2022} We employed a Monkhorst-Pack k-point mesh with dimensions $10 \times 10 \times 1$ and a plane wave energy cutoff of 300 Ry. 

\subsection*{Calculating Topological Invariants}

To calculate the Chern numbers for monolayer WTe$_2$ and the bilayer heterostructure, subsets of DFT single-particle Bloch functions were bijectively rotated onto sets of maximally-localized Wannier functions (MLWF's)\cite{Marzari_prb1997} starting from selected columns of the density matrix from DFT via the SCDM-k method.\cite{Damle_jcp2017} This mapping was performed for the isolated set of 31 monolayer 1$T'$-WTe$_2$ bands ranging from -10 eV below to 0.6 eV above the monolayer Fermi level, and for entangled sets of bilayer 1$T'$-WTe$_2$/CrI$_3$ bands as detailed in the Supplementary Information. All of the obtained MLWF's were well-localized and replicated the DFT band structure well over the span of bands involved in calculating topological invariants. Next, the hopping terms and correction terms for the lattice vectors of the hopping terms output by Wannier90 were used as input to the tight-binding model for the Z2Pack software for calculating topological invariants. Z2Pack is capable of calculating the evolution of hybrid Wannier charge centers across the surface defined by an explicit Hamiltonian $H(\textbf{k})$, a tight-binding model, or an explicit first-principles calculation.\cite{Soluyanov_prb2011,Gresch_prb2017} Thus, with our tight-binding model as input, we used Z2Pack to calculate the hybrid Wannier center evolution of the MLWF's corresponding to the bands up to and including the two orbitals involved in WTe$_2$'s spin-orbit-induced gap opening\cite{Qian_science2014} on a small k-space sphere with a radius of 0.001 centered at the $\Gamma$-point of the first Brillouin zone. All of the Z2Pack calculations passed the line and surface convergence checks to within the default tolerances of Z2Pack.\cite{Soluyanov_prb2011,Gresch_prb2017}

\subsection*{Simulating Edge Conductance}

Lastly, in order to obtain more direct evidence that our heterostructure behaves like a topological spin filter, we modeled the edge conductance of ML 1$T'$-WTe$_2$ by constructing a four-band $\textbf{k} \cdot \textbf{p}$ model, solving the corresponding scattering problem, and calculating the conductance of this model using the quantum transport software Kwant.\cite{Groth_njp2014} 

Our model consists of two regions: a semi-infinite conducting lead representative of BL 1$T'$-WTe$_2$/CrI$_3$ and a scattering region representative of ML 1$T'$-WTe$_2$, which are connected to one another. To mimic the magnetic field that would be induced by the CrI$_3$ in the bilayer portion, we subject half of the ML 1$T'$-WTe$_2$ to a magnetic field. The bilayer portion of the Hamiltonian ignores hybridization between WTe$_2$ and CrI$_3$ (see the Results Section for discussion). We provide the forms of the continuous Hamiltonians for the lead and scattering regions in the Supplementary Information.

Next, we discretized this model onto a rectangular grid with site spacing commensurate with ML WTe$_2$'s lattice parameters: $a = 3.50$ Å and  $b = 6.34$ Å. Both regions used previously reported $\textbf{k} \cdot \textbf{p}$ model parameters for pristine 1$T'$-WTe$_2$,\cite{Shi_prb2019} and the bilayer region contained an additional energy offset for the magnetic field-induced splitting of the WTe$_2$ bands. We then solved the scattering problem according to the Landauer-Büttiker formalism,\cite{Groth_njp2014} and applied current operators to the resulting eigenfunctions to calculate the conductance.

\section*{Results and Discussion \label{sec:results}}

\subsection*{Interlayer Charge Transfer and Magnetic Induction \label{sec:res-chmag}}

\begin{figure*}[ht]
  \setlength{\belowcaptionskip}{-18pt}
   \setlength{\abovecaptionskip}{2pt}
  \begin{center}
    \includegraphics[width = 6.0in]{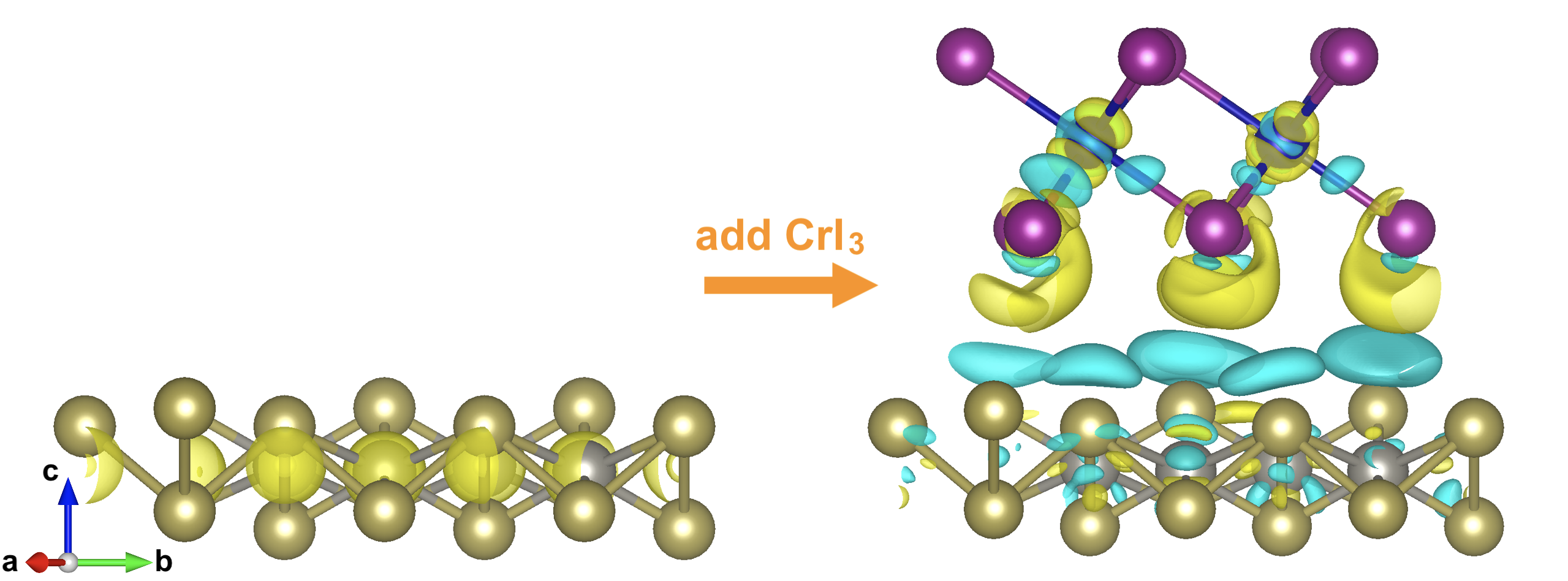}
  \caption{(Left) Collinear PBE charge density of monolayer WTe$_2$ in the absence of CrI$_3$ with an isosurface level value of 0.8. (Right) Collinear PBE+$U$ charge density difference between the bilayer and individual monolayers, $\rho_{BL}-\rho_{CrI3}-\rho_{WTe2}$, with an isosurface level value of 0.0025. Charge transfer from WTe$_2$ to CrI$_3$ is evident. Yellow indicates a positive charge density and light blue indicates a negative charge density. Chromium atoms are colored dark blue and iodine atoms are purple, while tungsten atoms are grey and tellurine atoms are beige.} 
  \label{fig3}
  \end{center}
\end{figure*}

As a first step toward understanding the physics of our bilayer, we began by examining how the layers influence each other's electronic structure. To do so, we analyzed the difference between the bilayer and individual monolayers' charge and spin densities. If proximity effects are truly at play, we would expect to see significant differences in their bilayer charge and spin densities relative to the separate monolayer densities. That said, when a monolayer of CrI$_3$ is stacked on a monolayer of 1$T'$-WTe$_2$, the DFT-predicted charge density difference between the bilayer and monolayers, $\rho_{BL}-\rho_{CrI3}-\rho_{WTe2}$, clearly shows charge accumulating near the CrI$_3$/WTe$_2$ interface as the charge is drawn downwards (see Figure \ref{fig3}). This suggests that CrI$_3$ is a charge acceptor and WTe$_2$ is a charge donor in the bilayer.  We see this charge transfer effect both with (non-collinear calculations) and without (collinear calculations) including SOC in our DFT calculations (see Figure \DSS{S3}), suggesting that it is a robust feature of the bilayer. 

Interestingly, charge also accumulates in between and around the bilayer with  charge accumulating near the CrI$_3$ within the vdW interface, and withdrawn from the portion of CrI$_3$ which is facing away from the interface. The existence of significant charge density within the bilayer gap confirms the strong hybridization between the iodine and tellurium atoms, which is also reflected in the significant atomic overlaps of all four atomic species in the partial densities of states or PDOS (see Figure \ref{fig5}). Additionally, the metallicity of the bilayer is reflected in the PDOS occupations of all four atomic species at and near the Fermi level, indicating that this hybridization causes CrI$_3$ to lose its Mott insulating nature when it is interfaced with WTe$_2$. The Lowdin charges of the constituent monolayers and bilayer are tabulated in Supplementary Tables \DSS{I} and \DSS{III} to quantify the extent of charge transfer in this bilayer. Summing the individual atomic charges of the monolayers and the bilayer yields an electron transfer of approximately 0.06 $e$ per primitive bilayer cell. 

\begin{figure}[ht!]
  \setlength{\belowcaptionskip}{0pt}
  \begin{center}
    \includegraphics[width = 1\linewidth]{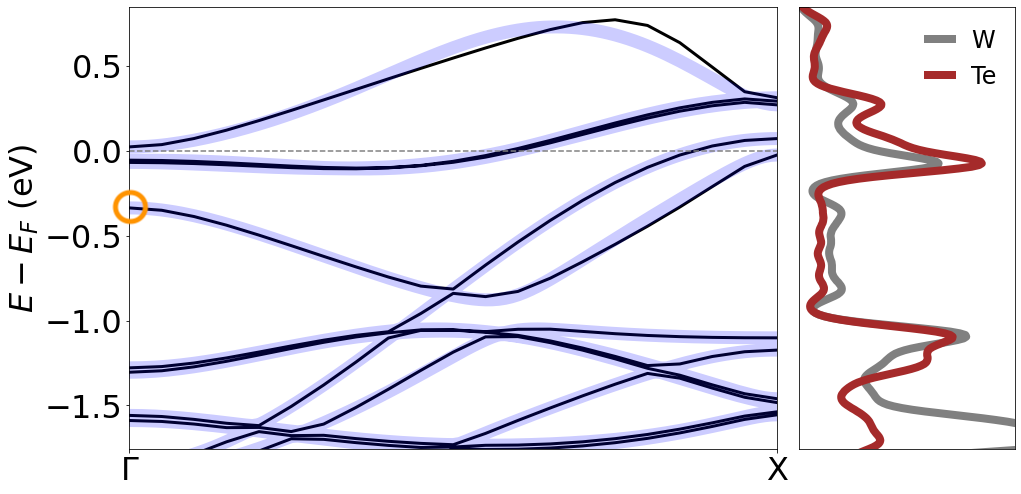}
    \includegraphics[width = 1\linewidth]{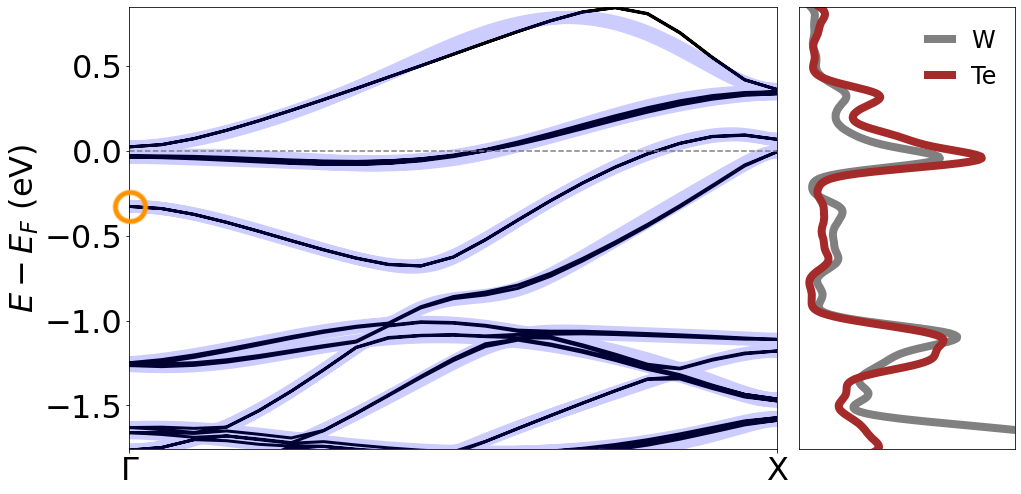}
  \caption{(Top) Collinear PBE band structure (black) and partial density of states (PDOS) of 0.8\%-strained 1$T'$-WTe$_2$ (relative to bulk) with the interpolated MLWF band structure overlain in light blue. (Bottom) Noncollinear PBE band structure with spin-orbit coupling.} 
  \label{fig4}
  \end{center}
\end{figure}

\begin{figure}[ht]
  \setlength{\belowcaptionskip}{0pt}
  \begin{center}
    \includegraphics[width = 1\linewidth]{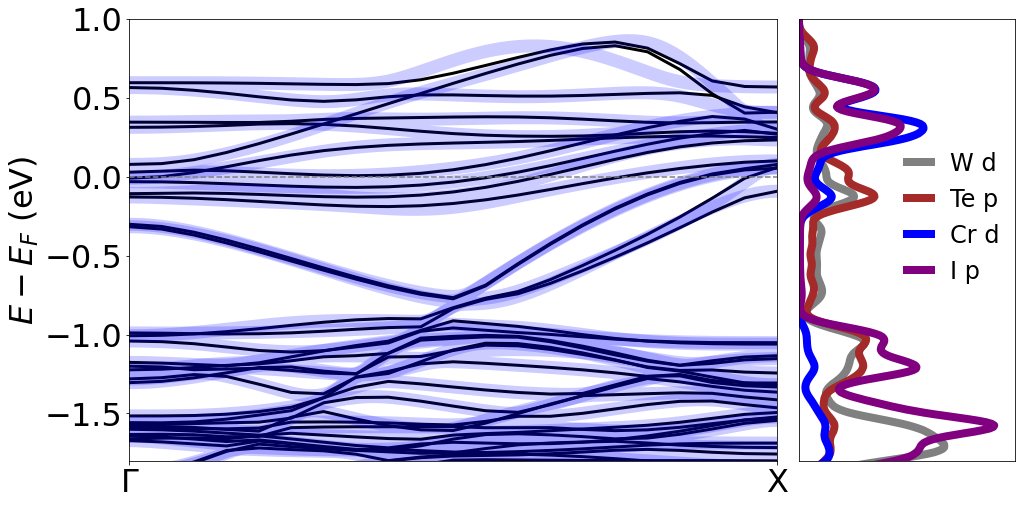}
    \includegraphics[width = 1\linewidth]{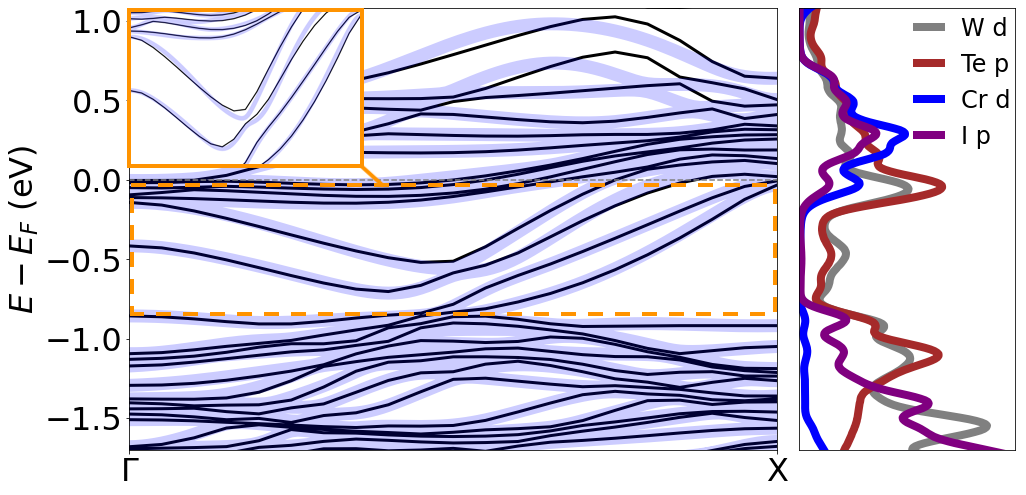}
  \caption{(Top) Collinear PBE+$U$ band structure (black) and partial density of states (PDOS) of bilayer 1$T'$-WTe$_2$/CrI$_3$ with the interpolated MLWF band structures overlain (light blue). (Bottom) Noncollinear PBE+$U$ band structure with spin-orbit coupling.} 
  \label{fig5}
  \end{center}
\end{figure}

Lastly, we consider the spin polarization that accompanies these charge transfer effects by evaluating the collinear spin density difference $s_{BL}-s_{CrI3}-s_{WTe2}$ and the $z$-component of the noncollinear magnetization density difference $m_{BL}-m_{CrI3}-m_{WTe2}$ between the bilayer and individual monolayers. We find that charge transfer in the absence of spin-orbit coupling (SOC) is minimally spin polarized, though the collinear spin density difference indicates that the Cr moments become slightly larger,  which can be attributed to charge transfer with a slightly larger spin-up character (see Figure \ref{fig2}). Strikingly, the charge transfer we observe is significantly more spin-polarized when SOC is considered. The charge transfer is mostly spin-up as in the collinear case, but to such a large extent 
that the W atoms in WTe$_2$ become polarized in the opposite, spin-down direction. We attribute this to the SOC-enhanced splitting of WTe$_2$’s majority up and down spin bands, which yields majority down spin WTe$_2$ bands that are lower in energy than its up spin bands similar to that observed in bilayer 1$T'$-WTe$_2$/CrBr$_3$.\cite{Bora_ass2023} This change is additionally accompanied by an increase in the magnitude of the bilayer Cr magnetic moments relative to those in monolayer CrI$_3$ with SOC (see Supplementary Table \DSS{IV}), which we attribute to the effect of strong Te/I hybridization on the anisotropic exchange interactions that stabilize monolayer CrI$_3$’s Ising-like ferromagnetism.\cite{Lado_2dmat2017} It is clear from the partial density of states in Figure \ref{fig5} that the ($e_g$) conduction bands of the bilayer with SOC have nearly equal Te and I $p$-orbital character, indicating strong hybridization. As the iodine SOC is instrumental in mediating the anisotropic exchange interactions which stabilize out-of-plane ferromagnetic order in monolayer CrI$_3$,\cite{Lado_2dmat2017} this hybridization must enhance the exchange interactions in a way that ultimately increases the magnitude of the $z$-component of the magnetic moments on the Cr atoms. Strong interlayer $e_g$-$e_g$ interactions in bilayer CrI$_3$ favor interlayer antiferromagnetic (AFM) coupling, \cite{Jang_prm2019} so the observed AFM interlayer coupling along with strong hybridization between the Te and I $p$-orbitals with $e_g$ character suggests a similar mechanism in this heterostructure.

\begin{figure*}[htp!]
  \begin{center}
    \includegraphics[width = 7.0 in]{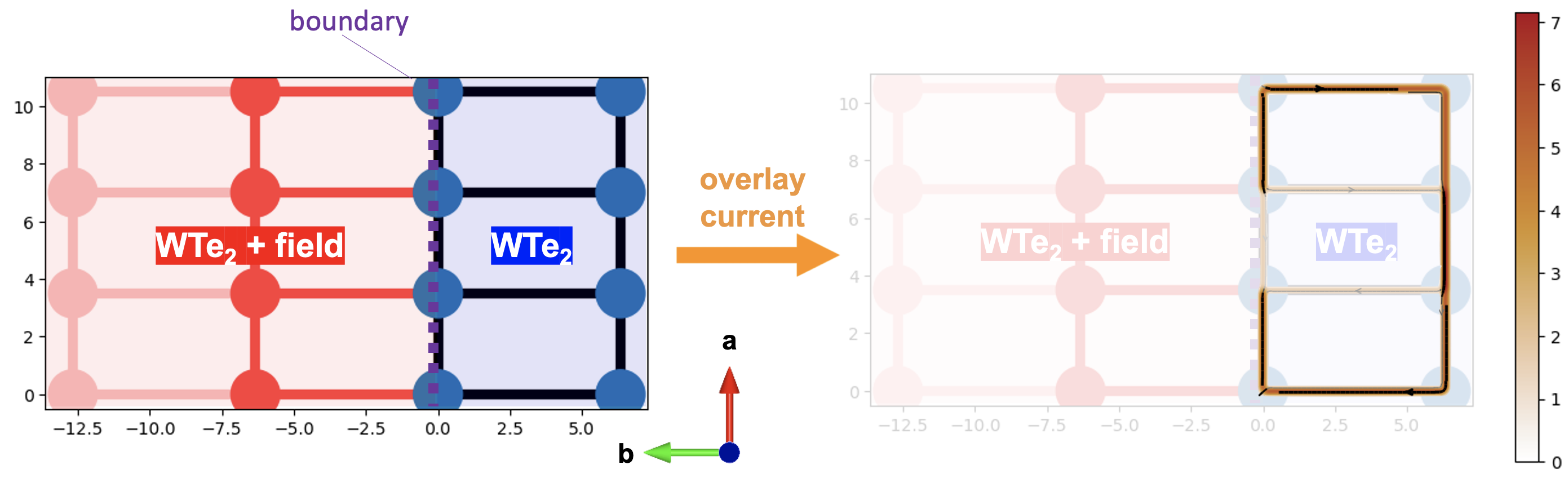}
  \caption{(Left) Top view of the sites in a simplified heterostructure in which the red region represents the lead corresponding to WTe$_2$ subject to the magnetic field produced by CrI$_3$ and the blue region is the scattering region of pristine ML WTe$_2$. (Right) The calculated current (in varying shades of red) overlain on a transparent version of the model heterostructure is largest around the edge of the pristine WTe$_2$.}
  \label{fig6}
  \end{center}
\end{figure*}

\subsection*{Topological Properties of ML 1$T'$-WTe$_2$ and BL WTe$_2$/CrI$_3$ \label{sec:res-chern}}

To assess the potential for topology that can give rise to chiral edge states in our bilayer, we first examined the band structures of the individual monolayers and combined bilayer structure. Previous modeling has shown that 1$T'$-WTe$_2$ possesses a band crossing below the Fermi level, which gives rise to its nontrivial topology.\cite{Tang_natphys2017} As a first step, we thus determined the band structures of our strained WTe$_2$ monolayer with and without spin-orbit coupling (SOC) to verify that our slight distortion does not change the bands significantly. 

Indeed, as shown in Figure \ref{fig4}, a band crossing occurs mid-way between the $\Gamma$ and X high-symmetry points as is also observed in simulations of pristine WTe$_2$ without spin-orbit coupling. Additionally, the density of states in this region has slightly more W $d$-orbital character than Te $p$-orbital character and is consistent with the previous assignment of these bands to the hybridized W 5$d_{xz}$ and 5$d_{z^2}$ orbitals.\cite{Tang_natphys2017} These band structures confirm that the strain applied to the monolayer did not alter its topology. Integration of the Berry curvature over the fiber bundle of MLWFs up to the band circled in Figure \ref{fig4} yields a Chern number of 0, meaning the topology is trivial as expected, since spin-orbit coupling was turned off. The same analysis of the noncollinear WTe$_2$ monolayer when spin-orbit coupling is included (bottom of Figure \ref{fig4}) exhibits a band gap within the 5$d_{xz}$ and 5$d_{z^2}$ bands about halfway between the $\Gamma$ and X high-symmetry points. Integration of the Berry curvature over the fiber bundle of these bands yields a Chern number of 1, verifying that this monolayer is topologically nontrivial.\cite{Tang_natphys2017}

Next, we calculated the band structure, partial density of states, and Chern number for the bilayer composed of ML 1$T'$-WTe$_2$ and ML CrI$_3$ (Figure \ref{fig5}) with and without SOC. This bilayer loses the Mott insulating behavior of CrI$_3$, with the bilayer exhibiting a metallic band structure and finite density of states at and around the Fermi level. The bilayer also maintains monolayer 1$T'$-WTe$_2$'s band crossing below the Fermi level, which opens when spin-orbit coupling is introduced as is visible in the upper panel of Figure \ref{fig5}. The integration of the Berry curvature over the fiber bundle of MLWF's up to and including the W 5$d_{xz}$ and 5$d_{z^2}$ orbitals yields a Chern number of 0, when spin-orbit coupling is not included in the DFT calculation, while the same process for the spin-orbit-coupled MLWF's also yields a Chern number of 0, indicating that this bilayer loses the topological character of monolayer 1$T'$-WTe$_2$ when CrI$_3$ is stacked on top of it. 

Since the constituent 1$T'$-WTe$_2$ layer is itself topological, this observation lends itself to the possibility of realizing a chiral edge state in a terraced 1$T'$-WTe$_2$/CrI$_3$ bilayer. According to the bulk-boundary correspondence,\cite{Hasan_rmp2010,Essin_prb2011} a chiral edge state should exist at the boundary between  topologically trivial and topologically nontrivial materials. If we place half of a layer of CrI$_3$ atop a layer of WTe$_2$ to form a terraced bilayer, the Chern number should change from 0 in the bilayer to 1 as soon as the Cr$I_3$ layer ends. This suggests that a chiral edge state should exist at this boundary. In addition, the strong hybridization of the WTe$_2$ with the ferromagnetic CrI$_3$ should break the degeneracy of the two chiral edge states of the ML WTe$_2$, causing the step edge of CrI$_3$ to host a spin-polarized chiral edge state. This suggests that a 1$T'$-WTe$_2$/CrI$_3$ terraced heterostructures could serve as a topological spin filter, as we attempt to further demonstrate below.

\subsection*{Conductance in a Model Terraced 1$T'$-WTe$_2$/CrI$_3$ Heterostructure \label{sec:res-edge}}
In principle, the bulk-boundary correspondence guarantees that a conducting edge must exist at the interface between topologically trivial and nontrivial regions of a system. In the spirit of completeness, we nevertheless performed a direct calculation of the conductance in a simplified model heterostructure which suggests that, even when interfacial scattering is considered, edge conductance persists at the boundary between monolayer WTe$_2$ and bilayer 1$T'$-WTe$_2$/CrI$_3$. 

The schematic in Figure \ref{fig6} depicts the hopping sites in our model heterostructure, as well as the calculated current, which is overlain onto a transparent version of the same grid. The red portion of the grid is a semi-infinite lead representing the conducting bilayer, from and to which current flows. This portion is represented by a four-band model which contains the four bands involved in the topological crossing of WTe$_2$\cite{Shi_prb2019} split by a magnetic field-induced potential that emulates the magnetic field produced by the proximate ferromagnetic monolayer CrI$_3$ (see Supplementary Information). The blue region is the scattering region of pristine monolayer 1$T'$-WTe$_2$, and it is represented by a four-band model which is the same as that of the lead except that it does not have the additional magnetic field-induced potential. Note that we do not include terms describing interlayer hybridization due to the fact that the primary effect of hybridization is to remove the topology of WTe$_2$, which was already established via our calculation of the Chern number in the bilayer portion of the terraced heterostructure. The length scale of our tight-binding model is set by the spacing of the grid points, which are equal to the lattice parameters of pristine ML 1$T'$-WTe$_2$ in units of Å.

Finally, we used the Kwant software package\cite{Groth_njp2014} to solve the scattering problem for the scattering matrix and eigenfunctions of our model heterostructure. We then constructed the current operator in terms of the scattering matrix at the energy at which the topological crossing occurs in the Kwant-computed band structure (see Supplementary Information), and calculated the conductance by applying this operator to the scattering eigenfunctions at that same energy and summing their contributions, obtaining a conductance of 7.0 $e^2/h$. When we plot the corresponding current in the right panel of Figure \ref{fig6}, it is evident that the current mostly flows around the edge, consistent with previous conductance experiments on pristine monolayer 1$T'$-WTe$_2$.\cite{Fei_natphys2017, Tang_natphys2017} In addition, this current is down-spin (polarized) since the proximity of CrI$_3$ lowers the energy of the down-spin bands in ML WTe$_2$.

\section*{Conclusions}
In conclusion, we have used a combination of Density Functional Theory and maximally-localized Wannier function-based tight-binding models to demonstrate that bilayer 1$T'$-WTe$_2$/CrI$_3$ is a topologically nontrivial metallic material which exhibits enhanced Cr magnetic moments and spin-polarized charge transfer from WTe$_2$ to CrI$_3$. Most notably, the topologically nontrivial monolayer WTe$_2$ becomes trivial when a monolayer of CrI$_3$ is placed on top of it, and this is reflected in a Chern number of 1 for monolayer 1$T'$-WTe$_2$ and of 0 for the bilayer. We further simulated the conduction of a simplified model of the terraced bilayer, finding that the edge conductance of WTe$_2$ is retained in such a heterostructure, and is spin polarized by the CrI$_3$. Taken together, our findings suggest that terraced bilayer 1$T'$-WTe$_2$/CrI$_3$ in which a monolayer of 1$T'$-WTe$_2$ is partly covered by a monolayer of CrI$_3$  may exhibit a chiral conducting edge state and is thus a candidate for being a topological spin filter. This is the first evidence for this type of behavior in a system composed of a $non$magnetic Weyl semimetal placed next to an atomically thin magnet, thus expanding the concept of such terraced chiral edge states beyond magnetic Weyl semimetal materials. 

\section*{Acknowledgements}

D.S. was funded by the U.S. Department of Energy through
the Office of Science Graduate Student Research (SCGSR)
Program. The funder played no role in study design, data collection, analysis and interpretation of data, or the writing of this manuscript.  This research was partly conducted as part of a user project at the Center for Nanophase Materials Sciences (CNMS), which is a US Department of Energy, Office of Science User Facility at Oak Ridge National Laboratory. P.G. and B.R. were supported by the U.S. Department of Energy, Office of Science, Basic Energy Sciences, Materials Sciences and Engineering Division, as part of the Computational
Materials Sciences Program and Center for Predictive Simulation of Functional Materials. This project used resources of the Oak Ridge Leadership Computing Facility under Contract No. DE-AC05-00OR22725 and of the National Energy Research Scientific Computing Center under Contract No. DE-AC0205CH11231, both U.S. Department of Energy Office of Science User Facilities. P.G. initiated and co-supervised the research. Simulations and analysis were performed by D.S.; the manuscript was prepared by D.S. and B.R. All authors read, edited, and approved the final manuscript. All authors declare no financial or non-financial competing interests.

\section*{Data availability}
The datasets used and/or analyzed during the current study are available from the corresponding author upon request.

\section*{References}

\bibliography{ref}

\makeatletter\@input{xx.tex}\makeatother
\end{document}


\include{MyCommand}
\title{Supplementary information for ``A first-principles study of bilayer 1T'-WTe$_2$/CrI$_3$ as a topological spin filter candidate"}

\author{Daniel Staros}
\affiliation{Department of Chemistry, Brown University, Providence, RI 02912}
\author{Brenda Rubenstein}
\email{Authors to whom correspondence should be addressed: Brenda Rubenstein, brenda\_rubenstein@brown.edu and Daniel Staros, daniel\_staros@brown.edu}
\affiliation{Department of Chemistry, Brown University, Providence, RI 02912}
\author{Panchapakesan Ganesh}
\affiliation{Center for Nanophase Materials Sciences Division, Oak Ridge National Laboratory, Oak Ridge, TN 37831, USA}

\maketitle
\date{\today}

\section{Calculating Topological Invariants: Wannier Function Subspaces} \label{supp-mlwf}
  \vspace*{-0.20cm}

In this section, we provide additional details about the selection of energy windows and bands used in the SCDM-k method\cite{Damle_jcp2017} for calculating the maximally localized Wannier functions (MLWF's)\cite{Marzari_prb1997} from which Chern invariants were calculated in Z2Pack.\cite{Soluyanov_prb2011,Gresch_prb2017} The Hilbert subspace used for calculation of MLWF's in monolayer $1T'$-WTe$_2$ is illustrated in Supplementary Figure \ref{figsi1}. The subspace is well-isolated from the other bands and the original subset of bands which were Wannierized is spanned by the obtained MLWF's (light blue).

\begin{figure}[hb!]
  \setlength{\belowcaptionskip}{4pt}
   \setlength{\abovecaptionskip}{16pt}
  \begin{center}
    \includegraphics[width = 1\linewidth]{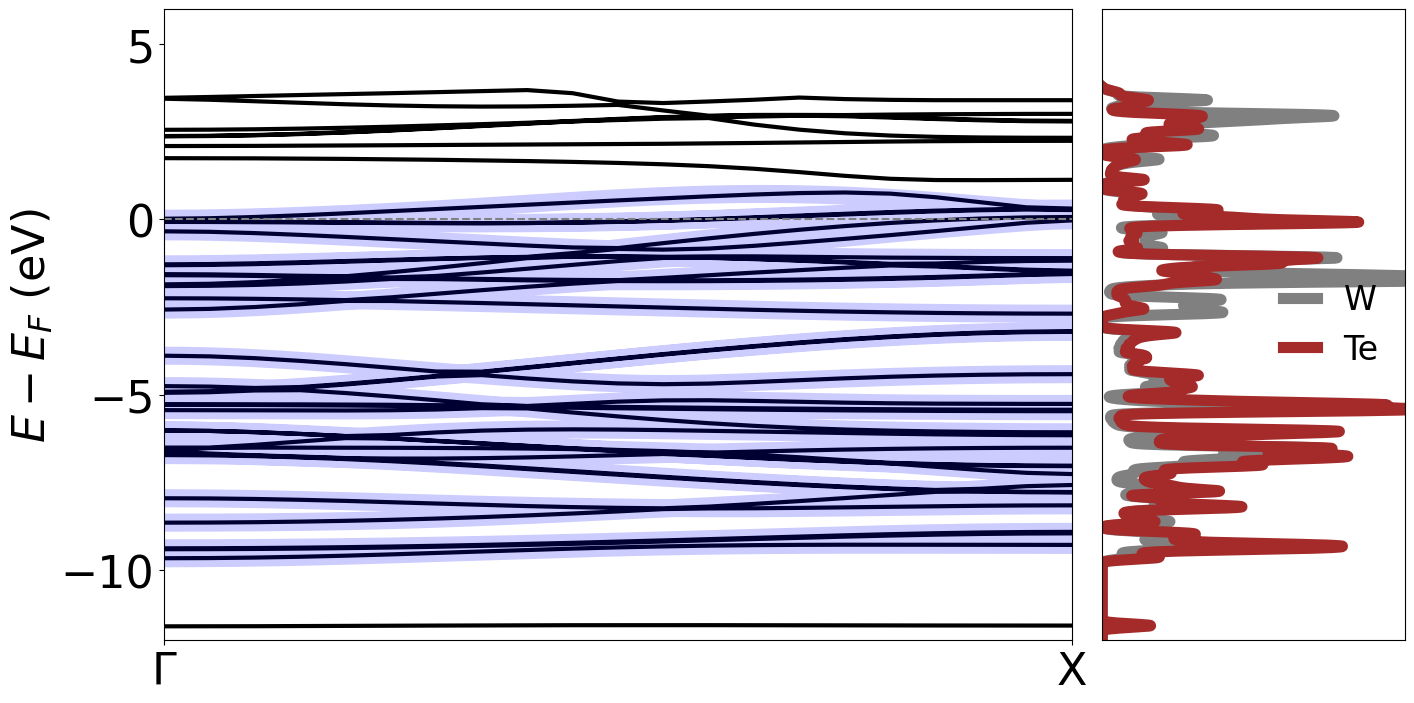}
    \includegraphics[width = 1\linewidth]{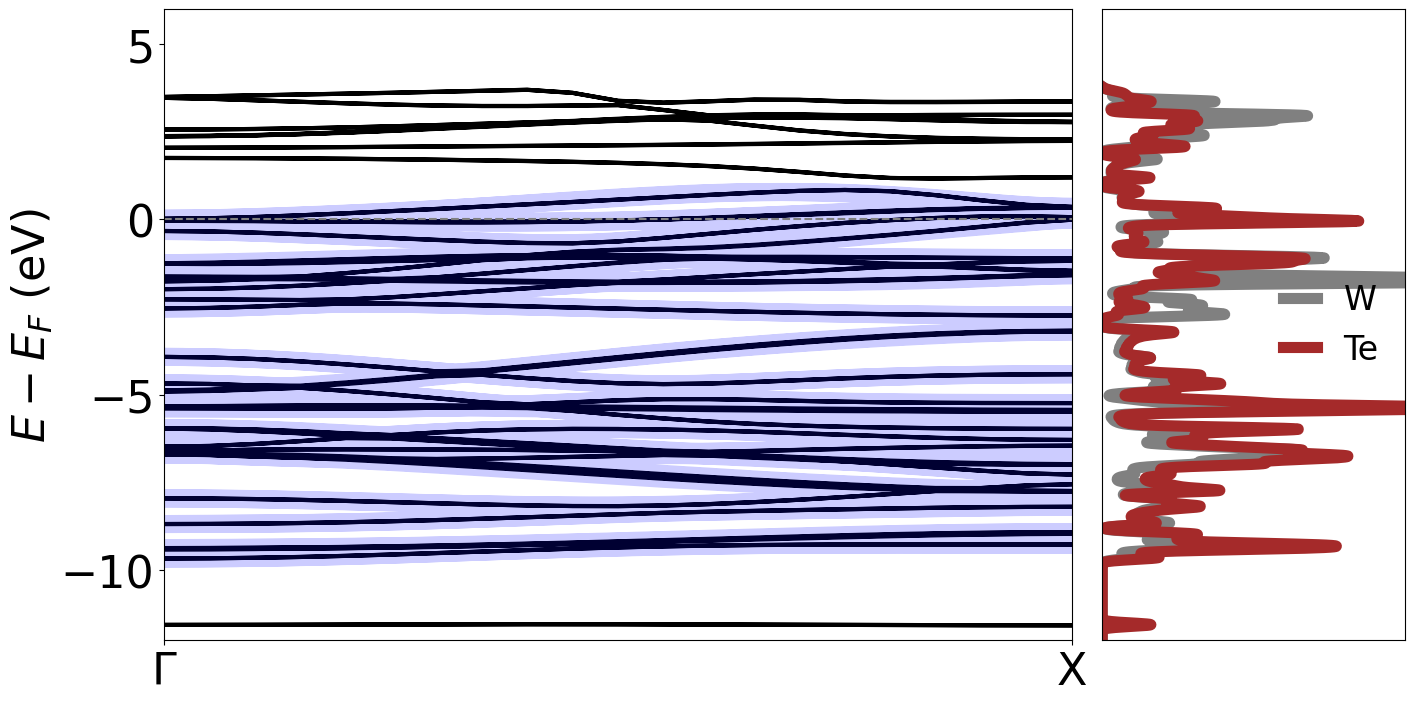}
  \caption{(Top) Zoomed-out collinear PBE band structure (black) and partial density of states of 0.8\%-strained 1$T'$-WTe$_2$ (relative to the bulk 1$T'$-WTe$_2$ monolayer) with the interpolated MLWF band structure overlain (light blue). (Bottom) Zoomed out noncollinear PBE band structure with spin-orbit coupling (black) and partial density of states of 0.8\%-strained monolayer 1$T'$-WTe$_2$ with the interpolated MLWF band structure overlain (light blue). The MLWF's are well-isolated and reproduce the chosen Hilbert subspace well.} 
  \label{figsi1}
  \end{center}
\end{figure}

\begin{figure}[hb]
  \setlength{\belowcaptionskip}{4pt}
   \setlength{\abovecaptionskip}{4pt}
  \begin{center}
    \includegraphics[width = 1\linewidth]{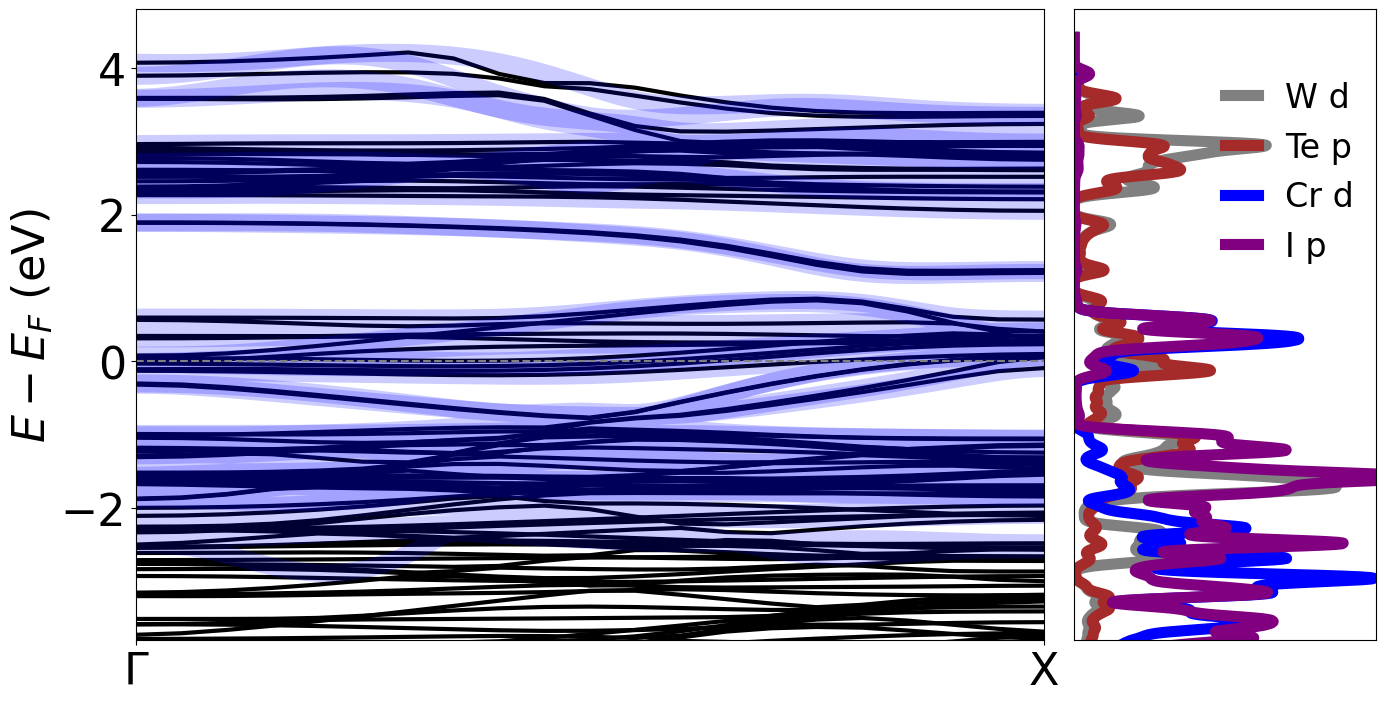}
    \includegraphics[width = 1\linewidth]{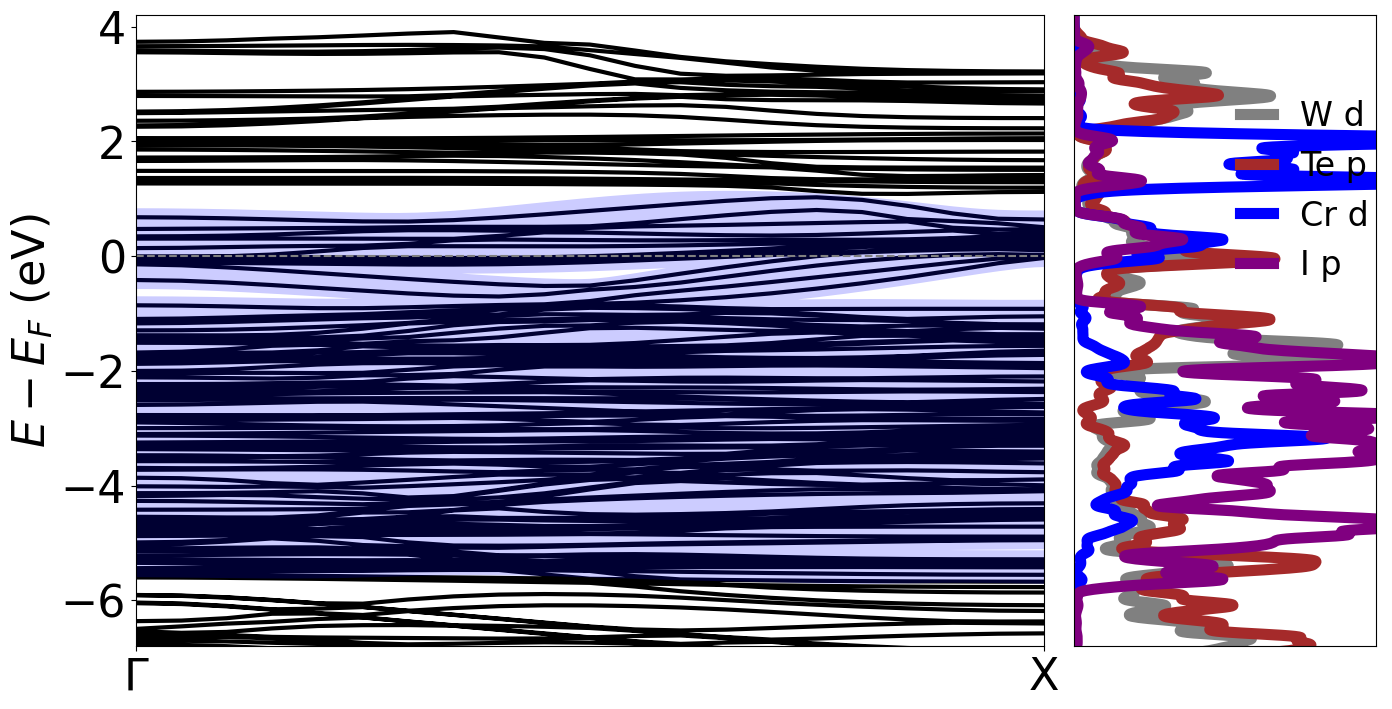}
  \caption{(Top) Zoomed out collinear PBE+$U$ band structure (black) and partial density of states of bilayer 1$T'$-WTe$_2$/CrI$_3$ with the interpolated MLWF band structures overlain (light blue). (Bottom) Zoomed out noncollinear PBE+$U$ band structure and partial density of states of bilayer 1$T'$-WTe$_2$/CrI$_3$ with the interpolated spinor MLWF band structure overlain. The MLWF's reproduce the chosen Hilbert subspace well around the Fermi level.} 
  \label{figsi2}
  \end{center}
\end{figure}

Additionally, the Hilbert subspace used for calculation of MLWF's in the bilayer required more careful evaluation. A fairly-small subspace for the collinear calculation resulted in MLWF's which were well-localized to within the dimensions of the cell, and their shape reliably reproduces that of the original bands over the relevant subspace (Supplementary Figure \ref{figsi2}). The bottom of the total subspace deviates slightly from the original bands due to the large entanglement of bands, but this does not affect calculation of the Chern number, which is determined solely by the bands in proximity to the Fermi level and not by the core bands, which are all topologically trivial. Additionally, both CrI$_3$ and WTe$_2$ in the absence of spin-orbit coupling are known already to be topologivally trivial. For the noncollinear MLWF calculations, a much larger band subspace was required to obtain robust MLWF's; these were localized to within the dimensions of the cell and reproduced the shape of the original bands over the entire Hilbert subspace despite being entangled at the bottom. 

\section{Interlayer charge transfer and magnetic induction: Atomic charges and moments} \label{charge-tran}
  \vspace*{-0.20cm}

\begin{figure}[h!]
  \setlength{\belowcaptionskip}{0pt}
   \setlength{\abovecaptionskip}{16pt}
  \begin{center}
    \includegraphics[width = 1\linewidth]{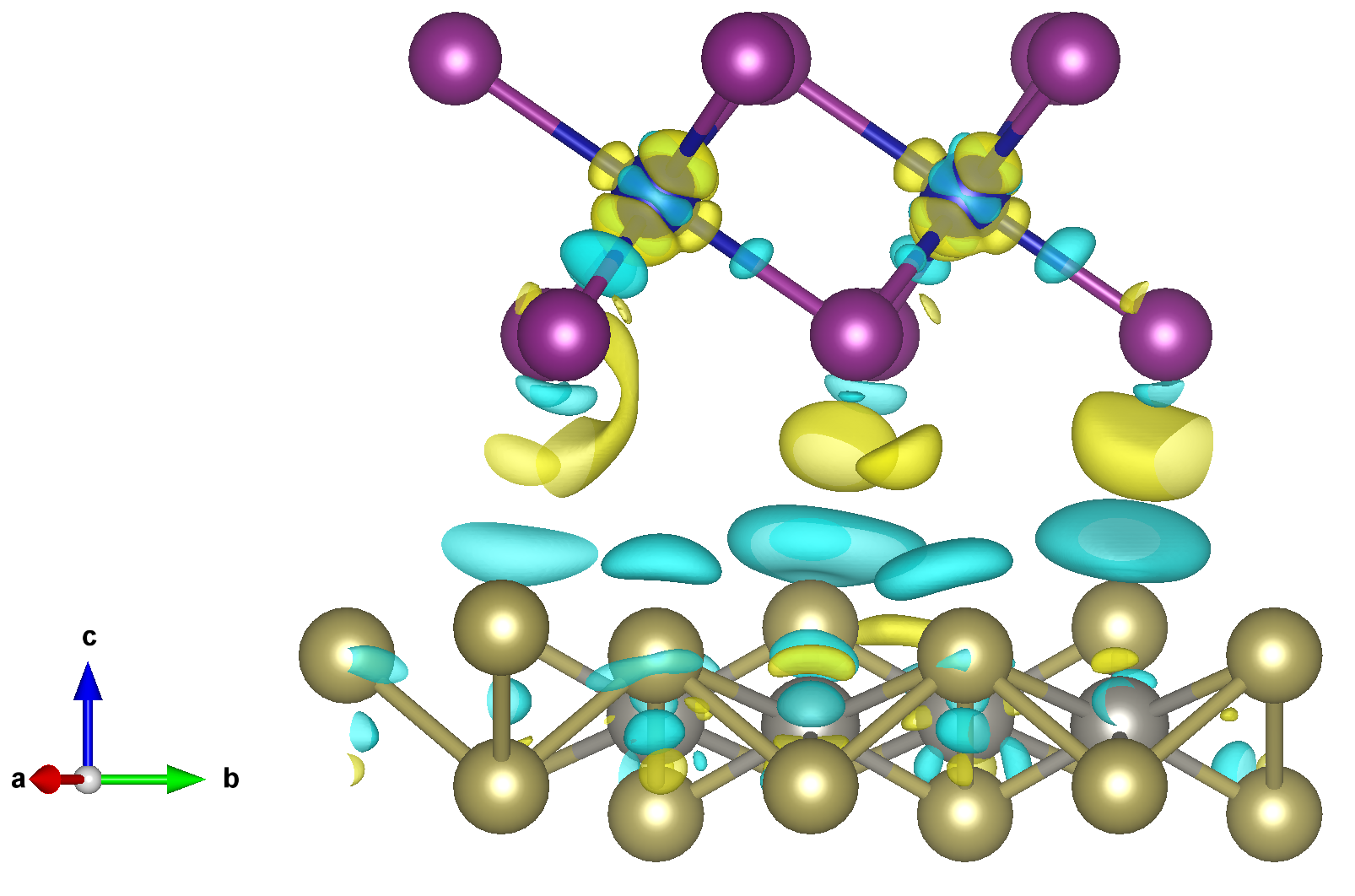}
  \caption{Noncollinear PBE+$U$ charge density difference between the bilayer and individual monolayers, $\rho_{BL}-\rho_{CrI3}-\rho_{WTe2}$ with an isosurface level value of 0.0025. Charge transfer from WTe$_2$ to CrI$_3$ is evident. Yellow indicates a positive charge density and light blue indicates a negative charge density. Chromium atoms are colored dark blue and iodine atoms are purple, while tungsten atoms are grey and tellurine atoms are beige.} 
  \label{figsi3}
  \end{center}
\end{figure}

Here, we further quantify the charge-transfer and magnetic induction in bilayer 1$T'$-WTe$_2$. Firstly, the charge-transfer from WTe$_2$ to CrI$_3$ is visible when spin-orbit coupling is included in the DFT calculations. As in the collinear case in the main manuscript, significant areas of negative charge density (blue) exist in the area between the two layers (Figure \ref{figsi3}). 

Additionally, we now tabulate the total atomic charges, magnetic moments, and individual orbital magnetic moments obtained from our DFT calculations. The total atomic charges and moments we report are those calculated automatically by Quantum Espresso (QE) at the end of self-consistent field (SCF) calculations using a semi-empirical weighted integration of electron or magnetization density around individual atomic spheres.\cite{Giannozzi_jpcm2009,Giannozzi_jpcm2017} Since Quantum Espresso SCF calculations do not provide orbital magnetic moments, we obtain these from the difference of the spin-polarized Löwdin charges reported in the partial density of states output files.

These data are separated into the collinear case (Tables \ref{tab:cl_charge} and \ref{tab:cl_mom}) and the noncollinear case (Tables \ref{tab:ncl_charge} and \ref{tab:ncl_mom}). As evidenced by changes only in the fourth significant digit of the atomic charges between collinear and noncollinear monolayers and bilayers, the inclusion of spin-orbit coupling changes atomic charges by very little. In the noncollinear bilayer, it is apparent that the Cr atoms gather a total of approximately 0.03 $e^{-}$ of charge and the I atoms gather a total of approximately 0.03 $e^{-}$ while the W atoms lose a total of approximately 0.06 $e^{-}$ compared to their constituent monolayers.

As the noncollinear calculation is more representative of the true physics of this system, we focus our discussion on the details of the noncollinear bilayer magnetic induction. Quantum Espresso reports the Cartesian components of magnetic moment $m_{atom}^x$, $m_{atom}^y$, and $m_{atom}^z$. Table \ref{tab:ncl_mom} contains the $m_{atom}^z$ values for individual atoms, and $m_{atom}^x$/$m_{atom}^y$ values were comparatively negligibly small, or in other words the Ising-like nature of CrI$_3$'s magnetism persists in both the noncollinear monolayer and the CrI$_3$-containing bilayer. 

When CrI$_3$ is placed on top of WTe$_2$, the magnetic moments on the Cr atoms are enhanced by about 0.3 $\mu_B$ from 2.9 to 3.2, the iodine magnetic moments are also enhanced by almost 0.1 $\mu_B$, and the W atoms gain atomic magnetic moments of about -0.1 $\mu_B$ which were not present at all in monolayer WTe$_2$ while the Te atoms gain negligible moments. These changes are an order of magnitude larger than the sum of changes in atomic charges, leading us to conclude that the magnetic induction effect is stronger in this bilayer than charge transfer, although both are present.

\section{Edge conductance model and scattering modes} \label{kwant-cond}
  \vspace*{-0.20cm}

The simplified heterostructure model used to simulate conduction consisted of continuous Hamiltonians, one for each region, discretized onto the grid shown in Figure 6 of the main text. Our effective four-band model for the blue scattering region was the same as the $\textbf{k} \cdot \textbf{p}$ Hamiltonian of pristine 1$T'$-WTe$_2$ given by Ref. \onlinecite{Shi_prb2019} and used the parameters given in their Appendix; it took the form
\begin{equation}
  \hat{H}_{Right} (\textbf{k}) = \bar{\varepsilon}_{\textbf{k}} I_4 +
  \begin{pmatrix}
m_{\textbf{k}}^{R} & v_{\textbf{k}}^+ & 0 & \; 0 \\
-v_{\textbf{k}}^- & -m_{\textbf{k}}^{R} & 0 & 0 \\
0 & 0 & m_{\textbf{k}}^{R} & v_{\textbf{k}}^- \\
0 & 0 & -v_{\textbf{k}}^+ & -m_{\textbf{k}}^{R}
  \end{pmatrix}
  \label{eq:ham_r}
\end{equation}

\noindent
where, we repeat from Ref. \onlinecite{Shi_prb2019}, $\bar{\varepsilon}_{\textbf{k}} = (\varepsilon_{c\textbf{k}}+\varepsilon_{v\textbf{k}})/2$, $I_4$ is the four-dimensional identity matrix $m_{\textbf{k}} = (\varepsilon_{c\textbf{k}}-\varepsilon_{v\textbf{k}})/2$ and $v_{\textbf{k}}^{\pm} = \pm v_x k_x + i v_y k_y$. The constituent terms may additionally be expressed in terms of the parameters provided by Ref. \onlinecite{Shi_prb2019} as follows: 
\begin{equation}
\begin{gathered}
    \varepsilon_{c\textbf{k}} = c_{1,0} + c_{1,x} k_x^2 + c_{1,y} k_y^2 \\ \varepsilon_{v\textbf{k}} = c_{2,0} + c_{2,x} k_x^2 + c_{2,y} k_y^2 \\ v_{\textbf{k}}^{\pm} = \pm v_{1,x} k_x + i v_{1,y} k_y
\end{gathered}
\label{eq:coeffs}
\end{equation}

\noindent
In order to describe the splitting of the WTe$_2$ bands by an external field for the bilayer portion of the heterostructure, we used a similar four-band Hamiltonian which captures the splitting of the bands by an external field potential $\Delta$ originating from the proximate CrI$_3$:
\begin{equation}
\begin{gathered}
  \hat{H}_{Left} (\textbf{k}) = \bar{\varepsilon}_{\textbf{k}} I_4 +
  \begin{pmatrix}
m_{\textbf{k}}^{L} + \Delta & v_{\textbf{k}}^+ & 0 & \; 0  \\
-v_{\textbf{k}}^- & -m_{\textbf{k}}^{L} + \Delta & 0 & 0 \\
0 & 0 & m_{\textbf{k}}^{L} & v_{\textbf{k}}^-  \\
0 & 0 & -v_{\textbf{k}}^+ & -m_{\textbf{k}}^{L}
  \end{pmatrix}
  \label{eq:ham_l}
\end{gathered}
\end{equation}

\noindent
where the definitions of Equation \ref{eq:coeffs} hold, and $\Delta = 0.005$ eV. Using Kwant, we discretized these Hamiltonians onto the hopping grid shown in the main text, and it was for this system that the scattering problem was solved to determine the conduction within our heterostructure.\cite{Groth_njp2014}

\begin{figure}[h!]
  \setlength{\belowcaptionskip}{0pt}
   \setlength{\abovecaptionskip}{16pt}
  \begin{center}
    \includegraphics[width = 1\linewidth]{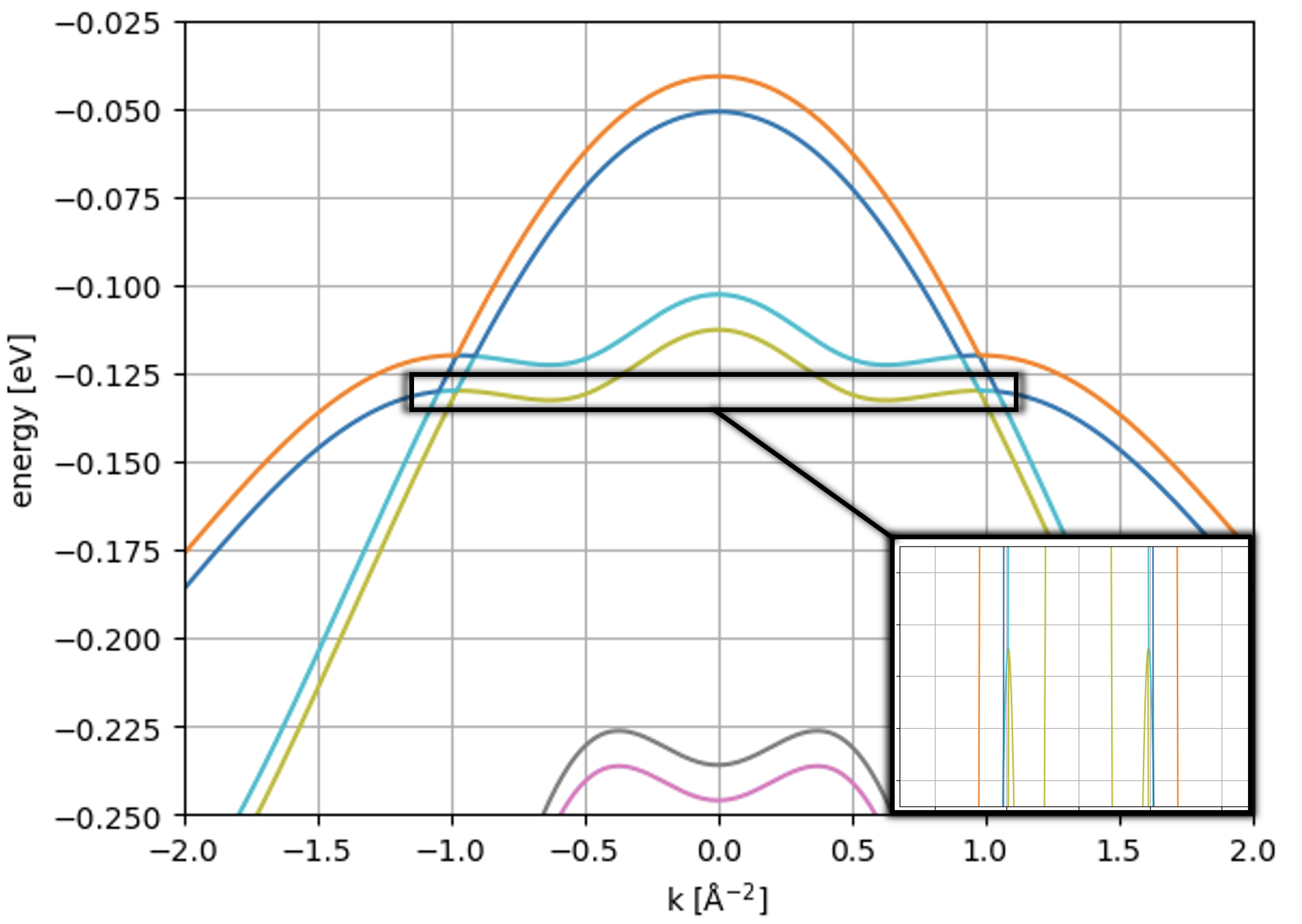}
  \caption{Kwant-generated band structure of the scattering eigenfunctions for the terraced heterostructure model. Note that several crossings appear due to the splitting of bands by our external potential. The black box indicates the crossing for which the conduction in the main text was calculated.} 
  \label{figsi4}
  \end{center}
\end{figure}

To determine the energy at which to evaluate the action of the current operator on the eigenfunctions of the scattering problem, we obtained a band stucture of the modes for this terraced system with Kwant (Figure \ref{figsi4}). Of the crossings which result, the one which occurs at -0.13 eV corresponds to the spin-split valence bands of WTe$_2$ and we highlight it in the inset of Figure \ref{figsi4}. Thus, we evaluated the current operator at this value, which yields the conduction plots presented in the main manuscript. 

\begin{table*}[hbt!]
\centering
\caption{Collinear DFT atomic charges in ML CrI$_3$, WTe$_2$ and BL 1T'-WTe$_2$/CrI$_3$}
\hspace*{-0cm}\begin{tabular}{||c c c c c c c c c c c c c c c c c c c c c ||}
\hline
& Cr1 & Cr2 & I1 & I2 & I3 & I4 & I5 & I6 & \; W1 \; \; & W2 \; & W3 \; \; & W4 \; & Te1 \; & Te2 \; & Te3 \; & Te4 \; & Te5 \; & Te6 \; & Te7 \; & Te8 \; \\ [0.5ex] 
 \hline\hline
\textbf{CrI$_3$} & 12.43 & 12.43 & 4.20 & 4.20 & 4.20 & 4.20 & 4.20 & 4.20 & - & - & - & - & - & - & - & - & - & - & - & - \\ 
\hline 
\textbf{ WTe$_2$} & - & - & - & - & - & - & - & - & 9.02 & 9.02 & 9.01 & 9.01 & 1.33 & 1.33 & 1.17 & 1.17 & 1.32 & 1.32 & 1.17 & 1.17 \\ 
\hline 
\textbf{BL} & 12.44 & 12.45 & 4.21 & 4.22 & 4.21 & 4.22 & 4.21 & 4.22 & 9.02 & 9.02 & 9.01 & 9.01 & 1.32 & 1.32 & 1.17 & 1.17 & 1.32 & 1.32 & 1.17 & 1.17 \\ 
\hline 
\end{tabular}
\label{tab:cl_charge}

\centering
\caption{Collinear DFT atomic moments in ML CrI$_3$, WTe$_2$ and BL 1T'-WTe$_2$/CrI$_3$}
\hspace*{-0cm}\begin{tabular}{||c c c c c c c c c c c c c c c c c c c c c ||}
\hline
& Cr1 & Cr2 & I1 & I2 & I3 & I4 & I5 & I6 & \; W1 \; & W2 \; & W3 \; & W4 \; & Te1 \; & Te2 \; & Te3 \; & Te4 \; & Te5 \; & Te6 \; & Te7 \; & Te8 \; \\ [0.5ex] 
 \hline\hline
\textbf{CrI$_3$} & 3.28 & 3.28 & -0.12 & -0.12 & -0.12 & -0.12 & -0.12 & -0.12 & - & - & - & - & - & - & - & - & - & - & - & - \\ 
\hline 
\textbf{ WTe$_2$} & - & - & - & - & - & - & - & - & 0.00 & 0.00 & 0.00 & 0.00 & 0.00 & 0.00 & 0.00 & 0.00 & 0.00 & 0.00 & 0.00 & 0.00 \\ 
\hline 
\textbf{BL} & 3.36 & 3.42 & -0.12 & -0.10 & -0.12 & -0.10 & -0.12 & -0.10 & 0.00 & 0.01 & 0.01 & 0.01 & 0.00 & 0.00 & 0.00 & 0.00 & 0.00 & 0.00 & 0.00 & 0.00 \\ 
\hline 
\end{tabular}
\label{tab:cl_mom}

\centering
\caption{Collinear ML CrI$_3$ atomic orbital moments}
\hspace*{-0cm}\begin{tabular}{||c c c c c c c c c ||}
\hline
& $d_{z^2}$ & $d_{xz}$ & $d_{yz}$ & $d_{x2y2}$ & $d_{xy}$ & $p_z$ & $p_x$ & $p_y$ \\ [0.5ex] 
 \hline\hline
\textbf{Cr1, Cr2} & 0.93 & 0.52 & 0.52 & 0.71 & 0.71 & 0.00 & 0.00 & 0.00 \\ 
\hline 
\textbf{I1, I2} & 0.01 & 0.00 & 0.01 & 0.01 & 0.01 & -0.07 & -0.02 & -0.11 \\ 
\hline 
\textbf{I3-I6} & 0.01 & 0.01 & 0.01 & 0.01 & 0.01 & -0.07 & -0.085 & -0.04 \\ 
\hline 
\end{tabular}
\label{tab:ml_orb_mom}

\centering
\caption{Collinear BL 1T'-WTe$_2$/CrI$_3$ atomic orbital moments}
\hspace*{-0cm}\begin{tabular}{||c c c c c c c c c ||}
\hline
& $d_{z^2}$ & $d_{xz}$ & $d_{yz}$ & $d_{x2y2}$ & $d_{xy}$ & $p_z$ & $p_x$ & $p_y$ \\ [0.5ex] 
 \hline\hline
\textbf{Cr1} & 0.93 & 0.54 & 0.57 & 0.73 & 0.72 & 0.00 & 0.00 & 0.00 \\ 
\hline 
\textbf{Cr2} & 0.93 & 0.60 & 0.58 & 0.75 & 0.72 & 0.00 & 0.00 & 0.00 \\ 
\hline 
\textbf{I1, I2} & 0.01 & 0.01 & 0.015 & 0.01 & 0.01 & -0.05 & -0.015 & -0.105 \\ 
\hline 
\textbf{I3-I6} & 0.01 & 0.015 & 0.01 & 0.01 & 0.01 & -0.06 & -0.078 & -0.043 \\ 
\hline 
\textbf{W1-W4} & 0.013 & 0.013 & 0.01 & 0.00 & 0.00 & 0.00 & 0.00 & 0.00 \\ 
\hline 
\textbf{Te1-Te6} & 0.00 & 0.00 & 0.00 & 0.00 & 0.00 & 0.01 & 0.00 & 0.00 \\ 
\hline 
\textbf{Te7,Te8} & 0.01 & 0.00 & 0.00 & 0.00 & 0.00 & 0.03 & 0.00 & 0.00 \\ 
\hline 
\end{tabular}
\label{tab:bl_orb_mom}

\centering
\caption{Noncollinear DFT atomic charges in ML CrI$_3$, WTe$_2$ and BL 1T'-WTe$_2$/CrI$_3$}
\hspace*{-0cm}\begin{tabular}{||c c c c c c c c c c c c c c c c c c c c c ||}
\hline
& Cr1 & Cr2 & I1 & I2 & I3 & I4 & I5 & I6 & \; W1 & \; W2 &  \; W3 & \; \; W4 & \; Te1 & \; Te2 & \; Te3 & \; Te4 & \; Te5 & \; Te6 & \; Te7 & \; Te8 \; \\ [0.5ex] 
 \hline\hline
\; \textbf{CrI$_3$ } \; & 12.43 & 12.43 & 4.20 & 4.20 & 4.20 & 4.20 & 4.20 & 4.20 & - & - & - & - & - & - & - & - & - & - & - & - \\ 
\hline 
\textbf{WTe$_2$} & - & - & - & - & - & - & - & - & 9.02 & 9.02 & 9.01 & 9.01 & 1.33 & 1.33 & 1.17 & 1.17 & 1.32 & 1.32 & 1.17 & 1.17 \\ 
\hline 
\textbf{BL} & 12.44 & 12.45 & 4.20 & 4.21 & 4.20 & 4.21 & 4.20 & 4.21 & 9.00 & 9.00 & 9.00 & 9.00 & 1.32 & 1.32 & 1.17 & 1.17 & 1.32 & 1.32 & 1.17 & 1.17 \\ 
\hline 
\end{tabular}
\label{tab:ncl_charge}

\centering
\caption{Noncollinear DFT atomic moment z-components in ML CrI$_3$, WTe$_2$ and BL 1T'-WTe$_2$/CrI$_3$}
\hspace*{-0cm}\begin{tabular}{||c c c c c c c c c c c c c c c c c c c c c ||}
\hline
& Cr1 & Cr2 & I1 & I2 & I3 & I4 & I5 & I6 & W1 & W2 & W3 & W4 & Te1 & Te2 & Te3 & Te4 & Te5 & Te6 & Te7 & Te8 \\ [0.5ex] 
 \hline\hline
\textbf{CrI$_3$} & 2.92 & 2.92 & -0.04 & -0.04 & -0.04 & -0.04 & -0.04 & -0.04 & - & - & - & - & - & - & - & - & - & - & - & - \\ 
\hline 
\textbf{WTe$_2$} & - & - & - & - & - & - & - & - & 0.00 & 0.00 & 0.00 & 0.00 & 0.00 & 0.00 & 0.00 & 0.00 & 0.00 & 0.00 & 0.00 & 0.00 \\ 
\hline 
\textbf{BL} & 3.20 & 3.24 & -0.13 & -0.12 & -0.13 & -0.13 & -0.13 & -0.12 & -0.08 & -0.08 & -0.07 & -0.08 & -0.01 & -0.01 & -0.01 & -0.01 & -0.01 & -0.01 & -0.01 & -0.01 \\ 
\hline 
\end{tabular}
\label{tab:ncl_mom}

\end{table*}

\section*{References}

\bibliography{ref}